\documentclass[10pt,final]{elsart}
\usepackage{amsmath,amssymb,graphicx}

\begin{document}
\begin{frontmatter}
\title{Quantum phases in artificial molecules\thanksref{si}}
\thanks[si]{To appear in Solid State Communications - Special Issue 
on Spin Effects in Mesoscopic Systems}
\author[modena]{Massimo Rontani\thanksref{email}},
\author[modena]{Filippo Troiani\thanksref{email2}},
\author[modena,graz]{Ulrich Hohenester}, and
\author[modena]{Elisa Molinari}
\address[modena]{%
  Istituto Nazionale per la Fisica della Materia (INFM) and
  Dipartimento di Fisica,\\
  Universit\`a degli Studi di Modena e Reggio Emilia,
  Via Campi 213/A, 41100 Modena, Italy}
\address[graz]{%
  Institut f\"ur Theoretische Physik,
  Karl--Franzenz--Universit\"at Graz, Universit\"atsplatz 5,
  8010 Graz, Austria  }
\thanks[email]{{\em Email address:} {\tt rontani@unimo.it}}
\thanks[email2]{{\em Email address:} {\tt troiani@unimo.it}}
\begin{abstract}
The many-body state of carriers confined in a quantum dot
is controlled by the balance between their kinetic energy
and their Coulomb correlation. In coupled quantum dots,
both can be tuned by varying the inter-dot tunneling and
interactions. 
Using a theoretical approach based on the diagonalization of 
the exact Hamiltonian, we show that transitions between 
different quantum phases can be induced through inter-dot 
coupling both for a system of few electrons (or holes) 
and for aggregates of electrons and holes. We discuss 
their manifestations in addition energy spectra (accessible
through capacitance or transport experiments) and 
optical spectra.
\end{abstract}
\begin{keyword}
A. semiconductors \sep
A. nanostructures \sep D. electron-electron interactions
\PACS 71.35.-y \sep 71.45.Gm \sep 73.23.-b \sep 73.23.Hk
\end{keyword}
\end{frontmatter}
\section{Introduction}
Semiconductor quantum dots (QDs) are a formidable laboratory for
next-generation devices and for the actual realization of 
some key {\em Gedankenexperimente} in many-body physics 
\cite{book1,book2,book3}. Indeed, the number of electrons and holes
in the QD can be controlled very accurately, and almost all 
relevant parameters influencing their strongly correlated states, 
like confinement potential and coupling with magnetic field and light, 
can be tailored in the experiments. 
The additional possibility of tuning the coupling between QDs enriches
their physics and the possible applications. 

From the point of view of fundamental physics such coupling extends
the analogy between quantum dots 
(``artificial atoms'' \cite{artificialatoms}) 
and natural atoms, to artificial and natural molecules.
The tunability of coupling among QDs allows to explore all regimes 
between non-interacting dots and their merging into a single QD; 
many of those regimes are precluded to molecular physics.

One of the peculiarities of QDs with respect to other solid state
structures consists in the partial decoupling of a few degrees of
freedom from all the others, which is due to the discrete nature
of the spectrum \cite{book1,book2,book3}.
The actual exploiting of such a feature largely
depends on the capability of integrating arrays of QDs, thus
increasing the number of degrees of freedom that one can address
with precision and coherent manipulation. This is precisely
the strategy pursued by the semiconductor-based 
solid state implementations
of quantum computation \cite{lloyd}.

In general and basic terms, the tuning of inter-dot tunneling
allows to modify the relative position of the single-particle levels,
thus inducing phase transitions in the many-body ground states and
different degrees of spatial correlations among carriers. 
Manifestations of these phenomena in systems formed by carriers 
of only one type, whose ground and excited state properties are accessible 
through addition energy spectra, have been predicted. 
Here we point out that similar effects are expected to occur also for 
systems formed by both electrons and holes. We also show that, in spite of
the obvious differences, strong similarities appear in the analysis
of electrons and electron-hole systems, and a unified theoretical 
description is in order.
Basically, a competition emerges between two trends. On one side
an atomic {\em Aufbau}\/ logic, where carriers tend to occupy the lowest
single-particle states available, thus minimizing the kinetic energy
and the total spin, at the (energetic) cost of reducing spatial
correlation among carriers. At the opposite extreme we find 
an enhanced degree of spatial correlation among carriers,
which occurs through the occupation of orbitals other than the lowest.
This implies an enhancement of the kinetic energy and a reduction of 
the repulsive one, and results in electron distributions maximizing the 
total spin (Hund's rule). The balance between these
two trends depends on the spacings of the single-particle levels involved,
and these are precisely what can be settled by controlling the
inter-dot tunneling. When carriers of opposite charge, different effective 
masses and tunneling parameters come into play, the competition between 
both trends becomes even more delicate.

Predictions of the actual ground and excited states of the 
many-body system thus require a careful theoretical treatment 
including all carrier-carrier interactions. Since the number of carriers
in the dot can be controlled 
and kept relatively small, we can proceed through 
direct diagonalization of the exact many-body Hamiltonian, with no need 
to make a priori assumptions on the interactions. On the contrary, the
results are a useful benchmark for the validity of the most common 
approximations for these systems.

We find that different quantum phases correspond to different regimes 
of inter-dot coupling both for a system of few electrons (or holes)
and for aggregates of electrons and holes, with various 
possible spatial configurations and the formation
of different possible ``subsystems'' of inter-correlated particles.
Besides, due to the negligible electron-hole exchange interaction
in heterostructures such as GaAs,
the two kinds of carriers can be treated as distinguishable particles.
Therefore spatial correlation among electrons and holes
does not arise from the Fermi statistics: it needs instead the
entanglement between the orbital degrees of freedom associated to holes
and electrons, and turns out to depend only indirectly 
on the spin quantum numbers $ S_{e} $ and $ S_{h} $.

After a brief summary of the state of the art in theoretical 
and experimental work on coupled dots (Sect.~\ref{Review}),
in the following we describe the general Hamiltonian and solution
scheme (Sect.~\ref{Method}). We then come to the results 
for electron- (Sect.~\ref{Electron}) and electron-hole systems 
(Sect.~\ref{Electron-hole}). The trends leading to different 
quantum phases are discussed in detail, together with their
nature in terms of spin and spatial correlation functions.

\section{Experimental and theoretical background}
\label{Review}

Early experimental and theoretical studies
focused on electrostatically-coupled dots with
negligible inter-dot tunneling \cite{review}.
Here we consider {\em artificial molecules}\/ \cite{leoscience},
where carriers tunnel at appreciable rates between dots, and
the wavefunction extends across the entire system.

The formation of a miniband structure in a one-dimensional array 
of tunnel-coupled dots was demonstrated more than a decade ago 
\cite{leocrystal}. After that, the first studies considered ``planar''
coupled dots defined by electrodes in a two dimensional electron gas.
In these devices the typical charging energy was much larger than the 
average inter-level spacing, hence 
linear \cite{planarlinear} and non-linear \cite{planarnonlinear}
Single Electron Tunneling Spectroscopy (SETS), obtained by transport
measurements at different values of the inter-dot conductance, could be 
explained by model theories based on capacitance parameterizations 
\cite{planarth}.
Early studies also considered 
simple model Hamiltonians (usually Hubbard-like)
with matrix elements treated as parameters \cite{Hubbardsimple}.
Blick and coworkers clearly showed the occurrence of coherent molecular
states across the entire two-dot setup,
analyzing transport data \cite{blickI} and the response to a
coherent wave interferometer \cite{blickII}. The tuning of coherent
states was also probed by microwave excitations \cite{PAT}, and
coupling with environment acoustic phonons was studied \cite{spontaneous}.
Planar coupled dots were also used to cool electron
degrees of freedom \cite{vaart}, to measure the magnetization
as a function of the magnetic field \cite{magnetization},
and to study the phenomenon of ``bunching'' of addition energies
in large quantum dots \cite{bunching}. 
The so-called ``vertical'' experimental geometry was introduced later:
it consists of a cylindrical mesa incorporating a triple barrier
structure that defines two dots. Sofar, evidence of single-particle
coherent states in a AlAs/GaAs heterostructure has been
reported \cite{schmidt}, while in 
AlGaAs/InGaAs structures clear SETS spectra
of few-particle states have been observed
as a function of the magnetic field $B$ and of the inter-dot barrier
thickness \cite{guy}.

A relevant part of theoretical research has addressed 
the study of few-particle states in vertical geometries, within
the framework of the envelope function approximation.
The two-electron problem was solved, by means of exact diagonalization,
in different geometries by Bryant \cite{bryant} and by Oh 
{\em et al.} \cite{oh}. Systems with a number of electrons $N>2$
at $B\approx 0$ in cylindrical geometry have been studied by several 
methods: Hartree-Fock \cite{tamura}, exact diagonalization for $N\le 5$ 
\cite{tokura}, numerical solution of a generalized Hubbard Model 
for $N\le 6$ \cite{ssc} and for $N>12$ 
with a ``core'' approximation suitable 
for the weak-coupling regime only \cite{asano}, density functional 
theory \cite{bart}.

Palacios and Hawrylak \cite{palacios} studied the energy spectrum
in strong magnetic field and negligible inter-dot tunneling with 
various methods ($N\le 6$), and established a connection between 
the correlated ground states of the double-dot system and those 
of Fractional Quantum Hall Effect systems in double layers.
In this perspective, Hu {\em et al.} \cite{dagotto} 
studied collective modes 
in mean-field theory, Imamura {\em et al.} \cite{aoki} 
exactly diagonalized 
the full Hamiltonian at strong $B$ and different values of tunneling 
($N\le 4$), Mart\'{\i}n-Moreno {\em et al.}\/ \cite{tejedor} considered 
the occurrence of canted ground states. Also the far-infrared response 
of many-electron states was analyzed with various techniques 
\cite{mayrock}. Another interesting issue is the relation between quantum 
and ``classical'' ground states \cite{peeters} as the radius of the dot 
is enlarged, when electrons arrange to minimize electrostatic repulsion 
because the kinetic energy is quenched \cite{yannouleas,helium}.

The electronic properties of planar dots have also been studied 
theoretically, through a variety of techniques: configuration interaction 
or analytical methods with various approximations \cite{eto,natalia}, 
or density functional theory for larger values of $N$ \cite{leburton}.
The infrared \cite{tapash} and the thermoelectric \cite{molenkamp}
response were considered. Systems of coupled QDs are also among the
most promising candidates for the implementation of semiconductor-based
quantum information processing devices: some of the current proposals
identify the qubits with either the spin \cite{qc} or the orbital degrees
of freedom associated to the conduction band electrons in QDs.

Research on few-electron systems in double dots is thus a new field 
in very rapid growth, with increasing focus on the possible quantum 
phases and how they can be driven by artificially controllable parameters 
such as inter-dot coupling, magnetic field, dot dimension. The study of 
such phases is expected to add insight into the physics of 
double layers, e.g.~the conditions for Wigner crystallization, 
and of strongly correlated systems in general. 
The amount of experimental data on many-body states
in artificial molecules is still limited, but the 
whole bunch of spectroscopic tools currently available 
(linear and non-linear transport, Raman spectroscopy) 
is now beginning to be employed (see e.g.~SETS spectra in 
the $B$-$N$ space, Ref.~\cite{Amaha} in this issue) 
and should allow the direct verification of theoretical 
predictions and a more general understanding of the basic phenomena and 
trends.

Also the optical properties of coupled QDs depend both
on the confinement of electrons and holes and on the effects 
of correlation among these carriers. In spite of their importance, however, 
such correlation effects are still largely unknown. 
From the experimental point of view, cleaved-edge overgrown samples
have been used \cite{schedelbeck:97}, but self-organized quantum dots 
are most commonly employed for optics. Their stacking was 
demonstrated \cite{fafard:00}, and the splitting of the excitonic
ground state in a single artificial molecule was studied as a function
of the inter-dot distance. The lines in the photoluminescence spectra
were explained in terms of transitions among excitonic states obtained by
single-particle filling of delocalized bonding and anti-bonding electron 
and hole states \cite{schedelbeck:97,bayer:01}. When a few photoexcited 
particles are present, however, the correlations induced by the 
carrier-carrier Coulomb interactions play a crucial role \cite{insertio};
single-particle tunneling and kinetic
energies are also affected by the different energetic spacings of
electron and hole single-particle states.
The correlated ground and excited states will thus
be governed by the competition of these effects, not included in previous
theoretical descriptions of photoexcited artificial molecules
\cite{filippoprl}.
A detailed understanding of exciton and multiexciton states in coupled
semiconductor QDs, however, is of great interest for the development of
the optical implementations of quantum-information processing schemes,
starting from the identification of well characterized qubits 
\cite{bennet:00}.  The possibility of complete optical control 
over the computational space formed by interacting excitons in quantum 
dots has recently been demonstrated in Refs.~\cite{troiani:00} 
and \cite{biolatti:00}.  We therefore expect that a systematic 
investigation of trends in the many-body phases of coupled dots will 
be actively pursued for systems of few electrons, and extended to
systems of photoexcited electrons and holes in the near future.

\section{Many-body states of $N$ electrons and holes}
\label{Method}

In the following we focus on the motion of few electrons 
and holes confined in two coupled quantum dots.
Our primary interest is in the correlated
nature of ground and excited states of the interacting system.

Hereafter we consider a simplified model where, within the 
envelope function and effective mass approximations,
two coupled identical
vertical dots are described by a separable confining potential
$V(\varrho,z)=V(\varrho)+V(z)$, with
$V(\varrho)=\frac 1 2 m^*\omega_0^2\varrho^2$
an in-plane parabolic potential [$\vec{\varrho}=(x,y)$, $m^*$ is the
effective electron (hole) mass, $\omega_0$ the characteristic frequency]
and $V(z)$ a double square quantum well along the $z$
direction (see Fig.~\ref{scheme}). Each well (of width $L$ and barrier
potential height $V_0$) corresponds to a dot; 
the coupling between the two dots
is controlled either by varying the inter-dot distance $d$ (width of
the inter-dot barrier) or the height of the inter-dot potential barrier.
To vary $d$ implies to consider differently grown
devices. The full many-body Hamiltonian 
$\mathcal{H}$ (in zero magnetic field)
is the sum of the single-particle terms 
$H^{(0)}(\vec{r})=-\hbar^2\nabla^2/(2m^*)
+V(\varrho,z)$ and of the two-body Coulomb interaction terms:

\begin{equation}
{\mathcal{H}}=
\sum_{\xi=e,h}\sum_{i=1}^{N_{\xi}}\left[
H_{\xi}^{(0)}\!\left(\vec{r}_{\xi}^{\;(i)}\right)+
\sum_{j<i}\frac{e^2}{\kappa_r \left|
\vec{r}_{\xi}^{\;(i)}-\vec{r}_{\xi}^{\;(j)} \right|}\right]-
\sum_{i=1}^{N_e}\sum_{j=1}^{N_h}
\frac{e^2}{\kappa_r \left|
\vec{r}_{e}^{\;(i)}-\vec{r}_{h}^{\;(j)} \right|}.
\label{e:hmanybody}
\end{equation}
Here $\kappa_r$ is the dielectric 
constant of the semiconductor medium,
and the subscript $e$ ($h$) refers to electrons (holes). Effective masses,
characteristic frequencies, and details of the double well
entering $H^{(0)}$ differ for electrons and holes.

We choose this geometry for two reasons: firstly, experimental devices
whose behavior can be described by this model are
currently studied by several groups, allowing for precise
tailoring of the dot geometry, strong spatial confinement, and hence
observation of spectral features beyond the simple Coulomb
Blockade behavior (e.g.~in SETS spectra). Secondly, the cylindrical 
vertical geometry,
contrary to in-plane devices, has the richest degree of symmetry, which
is particularly helpful to theoretical
work both in reducing the size of Hilbert space sectors and
in analyzing electronic configurations.
Specifically, $\mathcal{H}$ is invariant 
under any rotation in the spin space
(the total spin $S$ and its projection $S_z$ are therefore
conserved), rotation around the $z$ axis in real space (conservation
of the $z$-component of the orbital angular momentum $M$), inversion
with respect to the geometrical 
center of the system (parity conservation).
In complete analogy with Molecular Physics \cite{slater} and
for each species of carriers we introduce a spectroscopic
notation to classify electronic terms, 
namely eigenstates of $\mathcal{H}$:
$^{2S+1}M_{g,u}$.
Here $g$ ($u$) stands for even (odd) parity and
$M$ takes the labels $\Sigma$, $\Pi$, $\Delta$, $\ldots$
standing for $M=0,1,2,\ldots$
Actually, a $\Sigma$ term is also invariant under reflection
with respect to a plane passing through the symmetry axis:
in this case the notation takes the form
$^{2S+1}\Sigma^{\pm}_{g,u}$, where $\pm$ labels the sign change under
reflection \cite{insertioII}.

We are interested here in the evolution of the ground and excited states
as the inter-dot distance $d$ is varied. This feature shows a remarkable
difference between artificial and natural molecules: in the latter the
inter-nuclear distance is almost fixed, controlled by the nature of bonding,
while in the former it can be tuned by adjusting electrods or by
growing different sample devices.
Ground and excited states can be probed by several kinds of spectroscopies.
Theoretically, once the energy spectrum is known after numerical
diagonalization of $\mathcal{H}$, it is quite easy to
compute the relevant observable quantities.

A considerable achievement has been obtained by transport
spectroscopies, like single-electron capacitance tunneling
spectroscopy \cite{capacitance} or SETS \cite{tarucha}
for the ground state,
or non-linear tunneling spectroscopy \cite{leoexcited} for
the excited states.
In a transport experiment the chemical potential
$\mu(N)$ of the double-dot is measured as the number of electrons $N$
is varied charging the system one electron by one.
In fact, from the experimental value $\mu(N)$
one can infer information about the ground state, being
$\mu\left(N\right)=E_0\left(N\right)-E_0\left(N-1\right)$,
with $E_0(N)$ the ground-state energy of 
the $N$-body system \cite{spectroscopy}.
Our theoretical strategy is straightforward:
we compute the ground state energies $E_0(N)$ at different values of $N$,
and from these the chemical potential $\mu$ to be compared
with the spectra.
Single-dot far-infrared spectroscopies \cite{book1,book2,book3}
are unsuitable to probe the relative motion of electrons and hence
their correlation, because light only couples to the
center-of-mass motion (generalized Kohn theorem) \cite{kohn}.
This is also true for a system of vertically coupled quantum dots
with cylindrical symmetry, as long as the in-plane
confinement potential (orthogonal to the symmetry axis, e.g.~the growth
direction) is parabolic and the polarization of light is in the same plane.
However, this limitation does not hold for two-photon processes
like Raman scattering, where density fluctuations
can excite collective modes of the interacting system \cite{Raman}.

Finally, optical spectroscopy allows the study of few-particle states
including electrons and holes. In the lowest order the
light-semiconductor coupling is associated either to the absorption of
a photon and to the promotion of an electron from the valence to the
conduction band or to the reversed process, which is accounted for by a
Hamiltonian of the form $-\mathbf{E \cdot P}$, where $ \mathbf{E}$ is
the electric field and $ \mathbf{P} $ the material polarisation
\cite{haug:93}.
Within the rotating-wave and dipole approximations the
luminescence spectrum for a QD initially prepared in state
$ | \lambda \rangle $ can be computed according to Fermi's golden rule:
\begin{equation}
L_{ \sigma } ( \omega ) \propto \sum_{ \lambda' }
\: \vert \, ( P_{\sigma} )_{ \lambda' , \lambda } \, \vert^{2} \;
\delta( E_{ \lambda } + \hbar\omega - E_{ \lambda' } );
\end{equation}
here $ (P_{\sigma})_{ \lambda , \lambda' } $ are the dipole matrix
elements
corresponding to the transition
between states $\lambda$ and $\lambda'$ 
(through removal of one electron-hole
pair) and the the creation of a photon with helicity $\sigma=\pm$.

\section{Few-electron system}
\label{Electron}

In this section we study the system of interacting
carriers of the same species, e.g.~electrons.
Let us start from the simplest case, that is the
two-electron molecule.
A theorem due to Wigner \cite{wignerth}
guarantees that the ground state is always
a singlet if time-reversal symmetry is preserved \cite{singlet}:
however, dramatic alterations of the energy spectrum and wavefunction are
driven by the inter-dot distance $d$ and the characteristic dot radius
$\ell_0=(\hbar/m^*\omega_0)^{1/2}$.

\begin{figure}
\centerline{\includegraphics[width=0.65\columnwidth]
{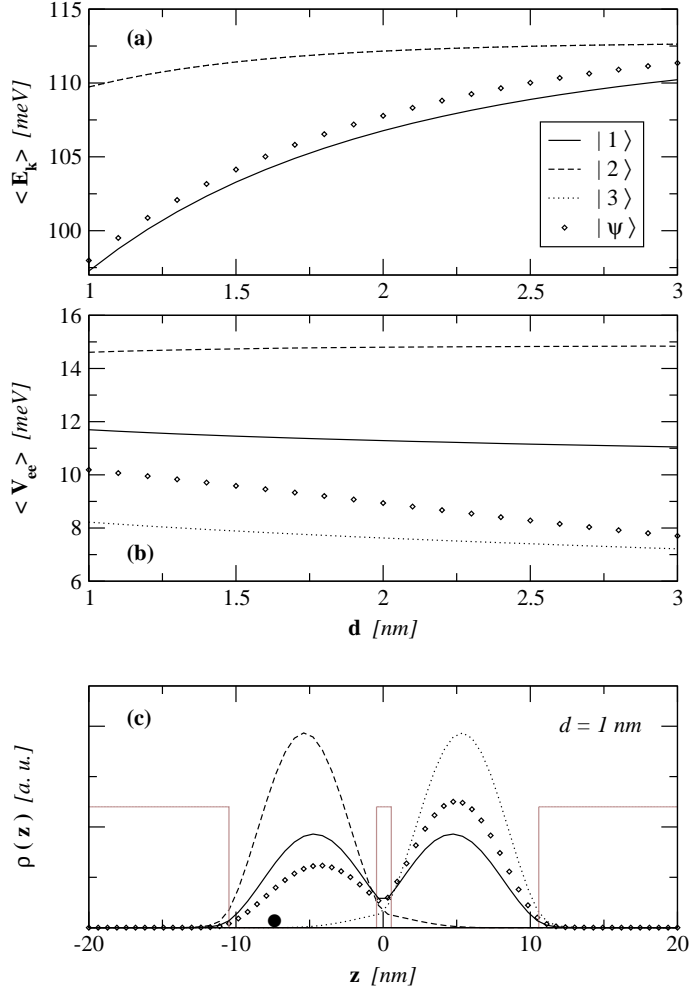}}\vspace*{0.8cm}
\caption{
Dependence on the inter-dot distance $d$ of: (a) the kinetic
$\left< E_k\right>$
and (b) Coulomb $\left< V_{ee} \right>$
energies of three prototypical two-electron states
$ | 1 \rangle \equiv | \sigma_{g}\! \uparrow ,
\sigma_{g}\! \downarrow \rangle $,
$ | 2 \rangle \equiv  ( | \sigma_{g} \!\uparrow , \sigma_{u} \!\downarrow
\rangle -  | \sigma_{g}\! \downarrow , \sigma_{u}
\!\uparrow \rangle ) / \sqrt{2}$
(singlet states) and
$ | 3 \rangle \equiv ( |\sigma_{g}\!\uparrow ,
\sigma_{u}\!\downarrow \rangle+
  | \sigma_{g}\! \downarrow , \sigma_{u}\! 
\uparrow \rangle ) / \sqrt{2} $
(triplet state); the
diamond symbol refers to the real ground state $ | \psi \rangle $.
In panel (c) we plot the spatially averaged 
pair-correlation function 
$ \rho ( z ) \equiv \int\!\int
P \:(x,y,z;x_{0},y_{0},z_{0}) \: {\rm d}x \, {\rm d}y $
corresponding to these
states. The coordinates of the fixed particle 
(represented by the black circle)
are $ x_{0} = 0 $ , $ y_{0} = 0 $ and $ z_{0} = -7.5 $ nm ( $ z = 0 $
is in the middle of the inter-dot barrier). 
The continuous grey line gives the
profile of the $z$ component of the 
confinement potential (barrier height
$ V_{0} = 400$ meV,
in-plane confinement energy $\hbar\omega_0=$ 20 meV).
Other parameters adopted are those typical of GaAs:
$ m_{e}^{*}= 0.067\: m_{e} $, $\kappa_r = 12.9$.
\label{scheme}
}
\end{figure}

This is shown in Fig.~\ref{scheme}: in panels (a) and (b) we plot
the total ground state kinetic
$\left< E_k\right>$ and Coulomb $\left< V_{ee} \right>$
energy \cite{insertioIII}, respectively, vs $d$.
Coulomb interaction mixes up different configurations (i.e.~Slater
determinants) which contribute with different weight to the ground state.
Besides $\left< E_k\right>$ and $\left< V_{ee} \right>$ for
the true few-particle groundstate
$|\psi\rangle$ (diamond symbol), 
in Fig.~\ref{scheme} we also show the corresponding data
of three prototypical states \cite{insertioIV}:
$ | 1 \rangle \equiv | \sigma_{g}\!\uparrow , 
\sigma_{g}\!\downarrow \rangle $
(singlet),
$ | 2 \rangle \equiv  ( | \sigma_{g}\! \uparrow ,
\sigma_{u}\! \downarrow \rangle -
  | \sigma_{g}\! \downarrow , \sigma_{u}\! 
\uparrow \rangle ) / \sqrt{2} $
 (singlet),
$ | 3 \rangle \equiv  ( | \sigma_{g}\! \uparrow ,
\sigma_{u}\! \downarrow \rangle +
  | \sigma_{g}\! \downarrow , \sigma_{u}\! 
\uparrow \rangle ) / \sqrt{2} $ (triplet).
Note that the difference between the (identical) kinetic energies of
states $ | 2 \rangle $ and $ | 3 \rangle $
and that of $ | 1 \rangle $ amounts
exactly to the energy splitting $\Delta_{sas}$
between the single-particle states
$ \sigma_{u} $ and $ \sigma_{g} $ [Fig.~\ref{scheme} (a)].
This quantity decreases exponentially as $d$ increases 
and as the probability of the tunneling
through the potential barrier goes to zero.
While singlet and triplet states
$ | 2 \rangle $ and $ | 3 \rangle $
have identical kinetic energy, the latter state is energetically favored
as the interaction energy is concerned. The splitting in
$\left< V_{ee} \right>$ between $ | 2 \rangle $ and $ | 3 \rangle $
appearing in Fig.~\ref{scheme} (b) is an {\em exchange energy,} namely the
consequence of the antisymmetry of the total wavefunction for
particle permutations.
The behavior of the ground state $|\psi\rangle$ partly
resembles that of the state $ | 1 \rangle$, but shows significant
deviations due to the mixing of configurations.

The arrangement of electrons is naturally visualized
by computing density functions in real space. However,
both the single-particle density and the usual
radial pair correlation function
$g(\varrho)$ plotted in the $xy$ plane depend only on the relative 
distance,
due to the cylindrical symmetry of the system. Hence, we follow
Ref.~\cite{maksym} and calculate the ``angular''
spin-resolved pair correlation function
\begin{equation}
P_{s,s_0}( \vec{\varrho} , z ;\vec{\varrho}_0 , z_0 )=A_{s,s_0}
\left<\sum_{i\ne j}\delta(\vec{\varrho}^{\;(i)}-\vec{\varrho})
\delta(z^{(i)}-z)
\delta_{s^{(i)},s}\delta(\vec{\varrho}^{\;(j)}-\vec{\varrho}_0)
\delta(z^{(j)}-z_0)\delta_{s^{(j)},s_0}\right>,
\end{equation}
where $\left<\ldots\right>$ denotes the expectation value on 
a given state,
the subscript $s$ refers to spin,
and $A_{s,s_0}$ is a normalization factor, such that
$\int {\rm d}\vec{\varrho}\,{\rm d}z
\,{\rm d}\vec{\varrho}_0\,{\rm d}z_0\, P_{s,s_0}$ $
( \vec{\varrho} , z ; \vec{\varrho}_0 , z_0 )=1$.
One electron with spin $s_0$ is fixed at the position $(\vec{\varrho}_0,
z_0)$, while the other at $(\vec{\varrho},z)$ with spin $s$ is varied: 
thus $P_{s,s_0}( \vec{\varrho} , z ; \vec{\varrho}_0 , z_0 )$ 
is proportional to the conditional probability of 
finding the second electron given that the first one is fixed. 
This allows for observation of the relative spatial
arrangement of electrons and of the angular correlation.
The spin-independent quantity is the total pair correlation function
$P( \vec{\varrho} , z ; \vec{\varrho}_0 ,
z_0 )$, normalized as
\begin{equation}
P( \vec{\varrho} , z ; \vec{\varrho}_0 , z_0 ) =\left[
\sum_{s=s_0}N_s(N_s-1)
P_{s,s_0} +
\sum_{s\ne s_0}N_sN_{s_0}
P_{s,s_0}\right]/N(N-1),
\end{equation}
being $N_s$ the number of spin-$s$ electrons.

In Fig.~\ref{scheme} (c)  we plot the function $ \rho ( z ) 
\equiv \int\!\int
{\rm d}x \, {\rm d}y \: P \:(x,y,z;x_{0},y_{0},z_{0}) $,
showing how the fixed position of one
electron (represented by the black circle) affects the spatial
distribution of the other one along the symmetry axis $z$, for an
inter-dot distance of 1 nm.
The state $ | 1 \rangle $
clearly exhibits no spatial correlation among
the two carriers: the placing of one electron in one quantum dot (QD)
does not change the
probability of finding the other one in any of the dots.
In the case of the singlet state $ | 2 \rangle $
the spatial distribution of one
electron is peaked around the other (fixed) one: the two
particles tend to occupy the same QD. Opposite trends apply
to the triplet state $ | 3 \rangle $.
Again, the true ground state shows a mixed character: 
$ \rho ( z )$ has its
biggest peak in the ``unoccupied'' QD, but there is a finite
probability for the double occupancy on the same QD.
The average values of the Coulomb energy
$\left< V_{ee}\right>$ [Fig.~\ref{scheme} (b)] clearly
reflect such behaviors. The curve referring to $ | 1 \rangle $
slowly decreases, because the value of $\left< V_{ee}\right>$
corresponding to two particles in different QDs 
diminishes as $d$ increases.
A fortiori, $ \langle
V_{ee} \rangle $ decreases for the triplet state: 
the electrons are always
in different QDs.
The Coulomb energy is less affected by the inter-dot distance
in the case of the state $ | 2 \rangle $,
because $ \langle V_{ee} \rangle $ is mainly
due to intra-dot interaction (both carriers in the same QD): the slight
increase of $ \langle V_{ee} \rangle $ depends on
the growing localisation of the particles within a QD.

The terms contributing to the Hamiltonian $\mathcal{H}$
of Eq.~(\ref{e:hmanybody}) scale
differently with the characteristic 
length of the confinement potential $\ell_0$:
the kinetic one goes like $\sim \ell_0^{-2}$,
while the interaction one like $\sim \ell_0^{-1}$.
For small dots, the kinetic term dominates and the system is
Fermi-liquid like: here the ground state is determined by the successive
filling of the empty lowest-energy single-particle levels, according
to {\em Aufbau} atomic theory. As $\ell_0$ increases, 
the electrons become
more and more correlated and arrange to minimize
Coulomb repulsion, up to the limit of complete spatial localization
(reminiscent of Wigner crystallization in 2D).
Even if for $N=2$ the ground state is always a $^1\Sigma_g$ term
as $\ell_0$ is varied,
nevertheless we can gain further insight into the 
correlation dynamics by
analyzing the two-body wavefunction.

\begin{figure}
\centerline{\includegraphics[width=0.85\columnwidth]{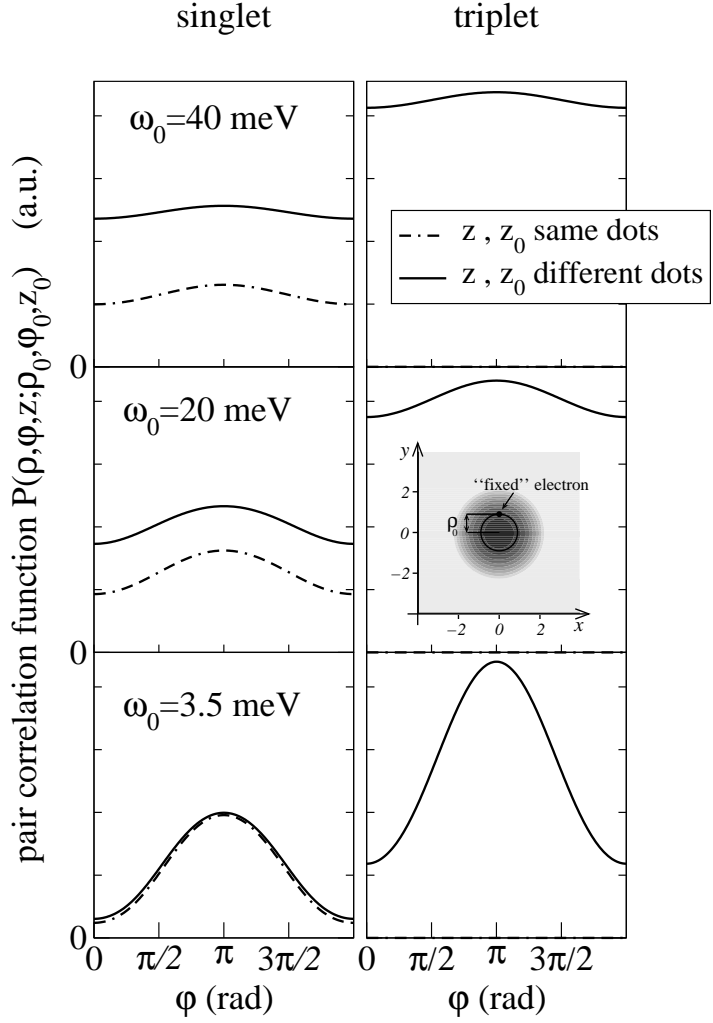}}
\caption{
Angular pair correlation function
$P( \varrho , \varphi , z ; \varrho_0 , \varphi_0 , z_0 )$
vs the azimuthal angle $\varphi$ for  $N=2$: 
all other parameters
$\varrho$, $\vec{\varrho}_0$, $z$, $z_0$ are fixed,
with $\varrho=\varrho_0$. Here $\varrho_0$, $z_0$, $z$ correspond
to the average value of the in-plane radius and the maxima along $z$
of the single-particle density, respectively.
We use $m^*=$ 0.067$m_e$, $\kappa_r=$12.9,
$L=$ 10 nm, $V_0=$ 400 meV, $d=$ 1 nm.
The inset is a contour plot of $P( \vec{\varrho} , z ; 
\vec{\varrho}_0 , z_0 )$
in the $xy$ plane (in units of $\ell_0$), for the triplet excited state
with $\hbar\omega_0=$ 20 meV.
\label{fthree}
}
\end{figure}

In the inset of Fig.~\ref{fthree} we plot the total pair correlation
function $P( \vec{\varrho} , z ; \vec{\varrho}_0 , z_0 )$ for the $N=2$
triplet state. Here, $\varrho_0$ and $z_0$ are set equal to
the average value of the in-plane radius and the maximum along $z$ of
the single-particle density, respectively. In addition, $z$ is fixed
at the position of the second, symmetric maximum of density in the
symmetry-axis direction:
the resulting contour plot is the value of $P$ in the $xy$ plane
(in units of $\ell_0$). This represents the probability of finding
one electron in the plane of one dot, given that the second electron
is fixed on the other dot.
The other plots in Fig.~\ref{fthree} show
$P( \varrho , \varphi , z ; \varrho_0 , \varphi_0 , z_0 )$
vs the azimuthal angle $\varphi$: all other parameters
$\varrho$, $\vec{\varrho}_0$, $z$, $z_0$ are fixed,
with $\varrho=\varrho_0$. When $\varphi=\varphi_0$, the position coincides
with that of the fixed electron, and the probability $P$ 
has a minimum (zero in the triplet case with $z=z_0$) due
to the Pauli exclusion principle. As $\varphi$ is varied, the position
follows a trajectory like the thick circle in the inset, starting from
the bullet locating the other fixed electron in the $xy$ plane.
After a $2\pi$-rotation, we are back in the starting point.
These plots are a kind of ``snapshot'' of the angular correlation,
as we freeze the motion of one electron. Figure \ref{fthree} is organized
in two columns, corresponding to the singlet ground state and to the
triplet excited state, respectively, for different values of
$\hbar\omega_0$ ($d=$ 1 nm).
Solid lines refers to the case $z=z_0$, namely electrons on the same dot,
while dashed lines to $z\ne z_0$, i.e.~electrons on different dots.
When $\hbar\omega_0$ is very large (40 meV, top row), the
curves $P( \vec{\varrho} , z ; \vec{\varrho}_0 , z_0 )$ are almost flat.
This flatness  implies that the motion 
of the two electrons is substantially
uncorrelated, except the effect of Fermi statistics. In fact,
in the triplet case, the probability of measuring two electrons
on the same dot is negligible, and this holds 
at any value of $\hbar\omega_0$.
On the contrary, in the singlet state there is a finite probability
of measuring two electrons on the same dot in the ground state.
As $\hbar\omega_0$ is reduced (20 meV, middle row), angular correlation
is turned on. This can be seen by the increase of the peak-valley ratio
in the angular correlation function. The position of the maximum
corresponds to $\pi$, i.e.~the two electrons repelling
each other tend to be separated as much as possible.
This trend is even clearer at small values of $\hbar\omega_0$ (3.5 meV,
bottom row). Plots of the second row ($\hbar\omega_0=$ 20 meV)
should be compared with the corresponding plot of Fig.~\ref{scheme} (c):
the first ones illustrate how electrons correlate in the $xy$ plane,
the second one how they arrange along $z$.

Note that in the first column of Fig.~\ref{fthree}
as $\hbar\omega_0$ is decreased
the probability of measuring two electrons on the same dot increases
up to the limit when it equals
the probability of measuring electrons on different dots
($\hbar\omega_0=$ 3.5 meV). This is due
to the different ratios between two fundamental energy scales: the
harmonic oscillator inter-level separation $\hbar\omega_0$ and the
energy difference $\Delta_{sas}$ between antisymmetric and
symmetric double-well wavefunctions: if $\hbar\omega_0\gg \Delta_{sas}$,
the inter-dot tunneling is negligible with respect to the kinetic energy
of the intra-dot motion, and the dots are almost quantum
mechanically decoupled. In the opposite limit, the system is coherent,
and it makes no difference between measuring one electron on one dot
or on the other one, since the system behaves as a unique
dot, doubled in size.

Let us now turn to discuss the case $N>2$. We choose a particular set of
parameters, namely $m^*=$ 0.067$m_e$, $\kappa_r=$ 12.4, $L=$ 12 nm,
$V_0=$ 250 meV, $\hbar\omega_0=$ 5.78$N^{-1/4}$, corresponding to a set
of experimental devices currently under study \cite{guy}. 
The parameterization
of  $\hbar\omega_0(N)$ is meant to mimic the effect of the gate voltage on
the electrostatic confinement potential $V(\varrho)$ \cite{bart}.
We exactly diagonalize the Hamiltonian $\mathcal{H}$ of
Eq.~(\ref{e:hmanybody}) for $N\le 6$, using up to 32 single-particle
orbitals. The convergence is checked controlling a cutoff on the average
energy of the Slater determinants entering the computation. Our code
uses the ARPACK package \cite{arpack} and isolates Hilbert
space sectors with $S$ and $S_z$ fixed, contrary to 
usual Lanczos approaches.

\begin{figure}
\centerline{\includegraphics[width=0.85\columnwidth]{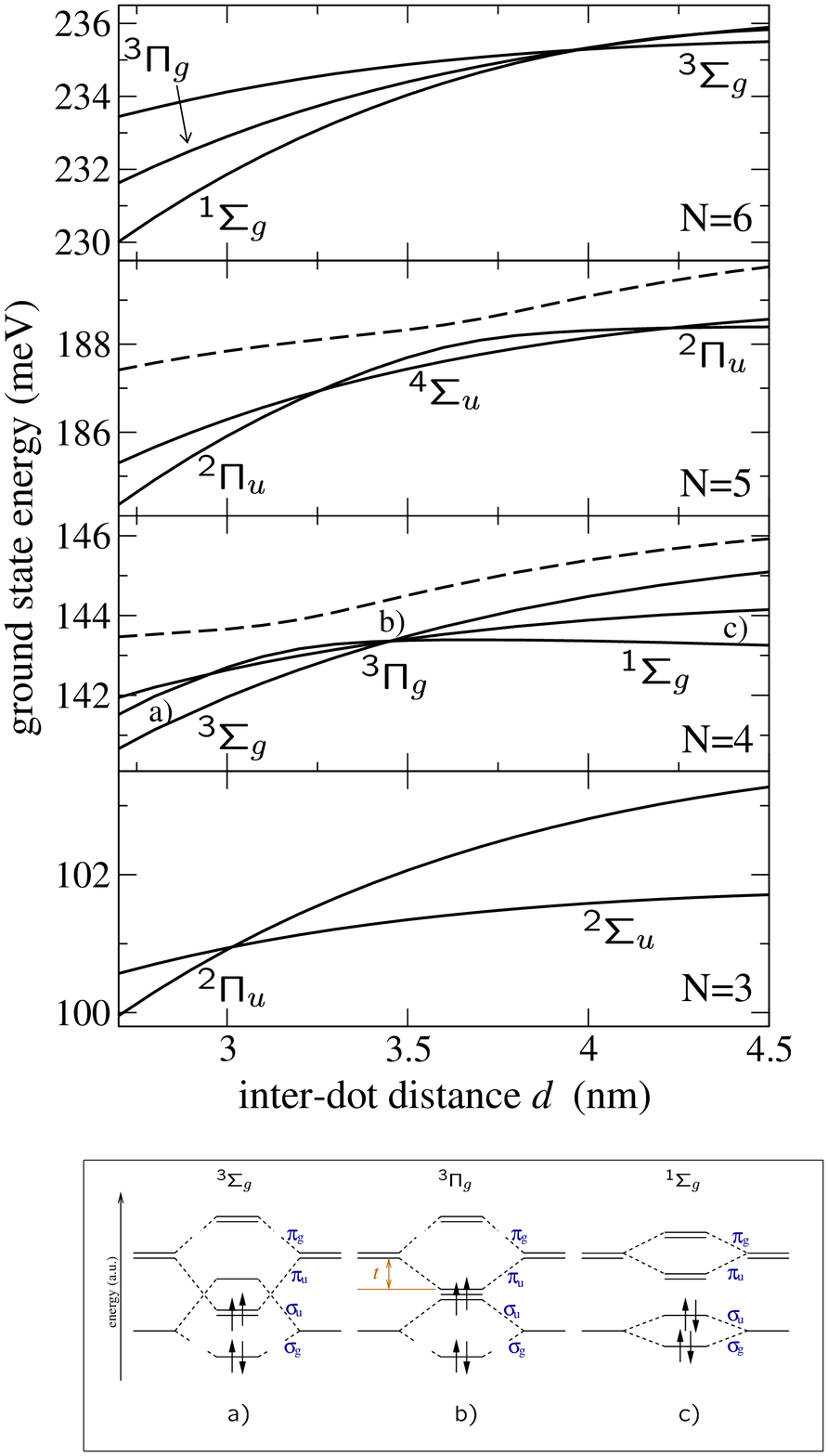}}
\caption{
Ground state energy vs $d$ for different number of electrons.
Some excited states are also depicted, toghether with their term in the
Molecular Spectroscopy notation. The bottom panel pictorially shows the
single-particle configurations that have the largest weight 
in the three different many-body ground states for $N=4$.
\label{f:four}
}
\end{figure}

Figure \ref{f:four} shows the calculated ground state energy vs $d$
for $3\le N \le 6$. As $d$ is varied, one or two transitions between
ground states of different symmetries occur. Specifically, while
there is only one transition between two different electronic terms for
$N=3$, two transitions take place for $N=4$ and $N=5$. 
The intermediate phase
for $N=4$ exists only in a very narrow range of $d$ ($\sim$ 0.01 nm)
in the neighborhood of $d=$ 3.45 nm. For $N=$ 6, 
again only two phases exist.
However, at the intersection point of the $^1\Sigma_g$ 
and $^3\Sigma_g$ terms,
the excited state $^3\Pi_g$ is almost degenerate in energy.

These transitions can be understood analyzing the many-body wavefunction
of the different ground states \cite{ssc}. In the bottom panel of
Fig.~\ref{f:four} we focus
on the $N=4$ case and we schematically depict the major-weight
Slater determinant corresponding to each phase. The key point is that,
as $d$ is decreased from the value of 4.5 nm,
the ``energy-gap'' $\Delta_{sas}$  between ``bonding'' and
``anti-bonding'' orbitals (i.e.~symmetric and antisymmetric
solutions of the double well along $z$) changes, from the limit of
decoupled dots [labeled as c) in figure]
to the strong-coupling limit [labeled as a)].
In the c) case, the first-shell molecular orbitals
$0\sigma_g$ and $0\sigma_u$ are 
almost degenerate and well separated in energy
with respect to the second shell, 
hence they are filled with four electrons
giving the configuration $0\sigma_g^20\sigma_u^2$, i.e.~two
isolated dots with the first orbital shell completely filled.
In the opposite limit, at small values of $d$ [a) case], 
the bonding mini-band
made of $0\sigma_g$, $0\pi_u^+$, $0\pi_u^-$ single-particle
orbitals is much lower in energy that the anti-bonding one.
The ambiguity of how to fill the lowest-energy orbitals,
due to the degeneracy of $0\pi_u^+$ and $0\pi_u^-$ levels, is solved 
consistently with 
Hund's first rule \cite{tarucha}, i.e. the two open-shell electrons
occupy each orbital with parallel spin (the configuration being
$0\sigma_g^20\pi_u^+0\pi_u^-$), in such a way that exchange
interaction prohibits electrons from getting close, minimizing
Coulomb repulsion. This configuration is
characteristic of a single dot, doubled in size \cite{tarucha}.
In the intermediate phase b), the antibonding $0\sigma_u$ level
is almost degenerate with the bonding $0\pi_u^+$ and $0\pi_u^-$ levels:
while the first two electrons occupy the lowest-energy orbital
$0\sigma_g$, the remaining two 
arrange again to maximize spin, consistently
to a ``generalized'' Hund's first rule. However, now there are
three levels almost degenerate, and we find that
the ground state configuration is $0\sigma_g^20\sigma_u0\pi_u^+$:
according to Hund's second rule, also the total orbital angular
momentum $M$ is maximized, to minimize the interaction energy
(the higher $m$, the smaller 
the Coulomb matrix element between single-particle
levels). A similar reasoning applies to the
transitions at $N\ne 4$.

Let us now focus on the $N=5$ case in Fig.~\ref{f:four}. As $d$ increases,
the ground state sequence is $^2\Pi_u \rightarrow\, ^4\Sigma_u
\rightarrow\, ^2\Pi_u$, that is the $^2\Pi_u$ term appears twice,
corresponding to a continuous energy curve that crosses twice
the $^4\Sigma_u$ term.
However, if we examine which Slater determinants mainly contribute 
to $^2\Pi_u$, we find that the relevant configuration at small $d$
(I $\equiv$ $0\sigma_g^20\pi_u^{+2}0\pi_u^-$) differs 
from that at large $d$
(II $\equiv$ $0\sigma_g^20\sigma_u^20\pi_u^+$).
Moreover, the slope of the curve in the
two regions is different, mainly due to the change in the balance between
bonding and anti-bonding levels occupied:  5:0 for I, 3:2 for II,
which controls the dependence of the overall kinetic energy on $d$
(see the previous discussion). This change in the ``character'' of the
$^2\Pi_u$ term (i.e.~the ratio between the weights
of configurations I and II) is found to be continuous with $d$.
We plot also the first excited state for the
$^2\Pi_u$ symmetry (dashed line in the $N=5$ panel): clearly this curve
anti-crosses the $^2\Pi_u$ ground state. Analyzing the slope and character
of this excited state in the small- and large-$d$ regions, we find
an inverted behaviour with respect to the ground state: now the relevant
configuration at small $d$ is II while that at large $d$ is I.
The overall behavior can be understood as a consequence of the
Wigner-von Neumann theorem, i.e.  that 
intersection of terms of identical symmetry is forbidden \cite{landau}.
Therefore, the two $^2\Pi_u$ terms anti-cross,
while $^2\Pi_u$ and $^4\Sigma_u$ terms
can freely cross and bring about ground state transitions,
belonging to different irreducible representations of
the symmetry group of $\mathcal{H}$. An analogous anti-crossing between
ground and excited state for the $^1\Sigma_g$ symmetry is depicted
in the $N=4$ panel (solid and dashed line, respectively).

Results of Fig.~\ref{f:four}
should be compared with those obtained by means of
exact diagonalization of a generalized Hubbard model \cite{ssc},
by density functional theory \cite{bart},
and by Hartree-Fock method \cite{tamura}. In all these
works the window in $d$-space at which the ground state $^3\Pi_g$
at $N=4$ occurs is much larger and a ghost additional intermediate phase
at $N=6$ (corresponding to the excited state $^3\Pi_g$ in our
calculation) appears.
Therefore our results, that agree well with data obtained up to $N\le 5$
by exact diagonalization in Ref.~\cite{tokura}, clearly demonstrate
the importance of correlation beyond mean-field approaches.
The interacting electronic system is so correlated in regimes
of realistic parameters of the devices that it is very
difficult to obtain quantitatively reliable results with
any approach but configuration interaction.
This point was already stressed, for single
quantum dots, in Refs.~\cite{pfannkuche} and \cite{prb}.

\begin{figure}
\centerline{\includegraphics[width=0.85\columnwidth,
angle=90]{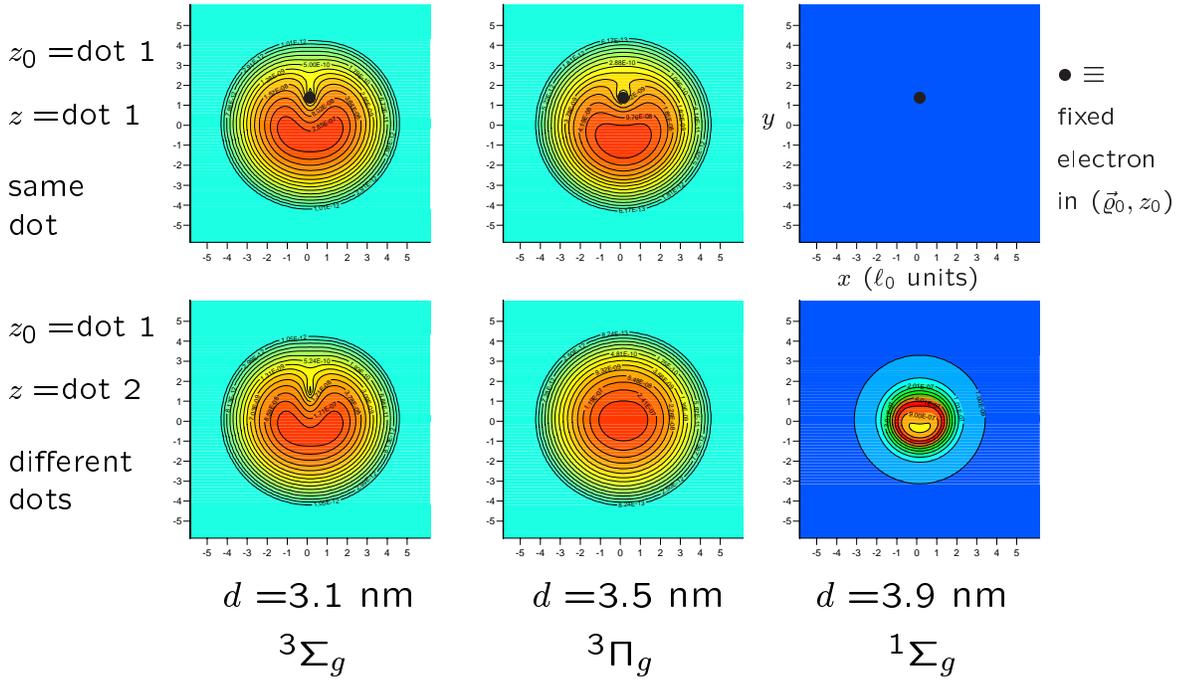}}
\caption{
Spin-resolved pair correlation function
$P_{\uparrow,\uparrow}( \vec{\varrho} , z ; \vec{\varrho}_0 , z_0 )$
for $N=4$ for three different values of $d$ (columns).
The position of a spin-up electron
$(\vec{\varrho}_0 , z_0 )$ is fixed in one dot as in Fig.~\ref{fthree}
(black bullet), and the contour plots of the top (bottom) row,
with $z=z_0$ ($z\ne z_0$) fixed,
correspond to the probability of measuring another spin-up electron
on the same (other) dot in the $xy$ plane ($\ell_0$ units).
\label{f:five}
}
\end{figure}

To further characterize different ground states vs $d$, 
in Fig.~\ref{f:five}
we plot the spin-resolved pair correlation function
$P_{\uparrow,\uparrow}( \vec{\varrho} , z ; \vec{\varrho}_0 , z_0 )$
for $N=4$ for values of $d$ corresponding to the three phases previously
discussed. The position of a spin-up electron
 $(\vec{\varrho}_0 , z_0 )$ is fixed in one dot as in Fig.~\ref{fthree},
and the contour plots of the top (bottom) row, 
with $z=z_0$ ($z\ne z_0$) fixed,
correspond to the probability of measuring another spin-up electron
on the same (other) dot in the $xy$ plane. The right column refers
to the $^1\Sigma_g$ term at $d=$ 3.9 nm: 
there is only one ``free'' spin-up electron, and we can see that
the probability of measuring it on the same dot as the fixed electron
is negligible,
while the probability distribution on the other dot
depends only slightly on the position of the first fixed electron.
Therefore the two dots are quantum mechanically
decoupled, each one filled with
two electrons in the lowest shell. 
The motion of electrons in the two dots
is almost uncorrelated in the $xy$ plane.
In the opposite limit of small $d$ (left column, $d=$ 3.1 nm, three
spin-up electrons), the coupling is so strong that it makes no difference
either measuring the electron on 
one dot or on the other, i.e.~the contour
plots are identical, and the system 
forms a coherent, strongly bound molecule.
The fixed electron is ``dressed'' by its exchange
and correlation hole, i.e.~it repels other electrons that are
at small distances. The middle column ($d=$ 3.5 nm, three
spin-up electrons) shows that in the intermediate phase $^3\Pi_g$
the dots are coupled with a weak degree of coherence,
namely the probabilities of measuring the electron on the two dots
are sensitively different. 
Planar correlation in the same dot is important,
while it is negligible for motion 
on different dots. The above classification
of phases is also suitable to $N\ne 4$.

In this section we have shown results for the electron energy spectrum
up to $N\le 6$. It is straightforward now to compute the SETS linear
spectrum, and the comparison with very recent experimental 
data \cite{Amaha} show remarkable agreement in many respects.
Results for $B > 0$ will be presented elsewhere \cite{tobepubl}.

\section{Few electron-hole pairs}
\label{Electron-hole}

\begin{figure}
\centerline{\includegraphics[width=0.85\columnwidth]{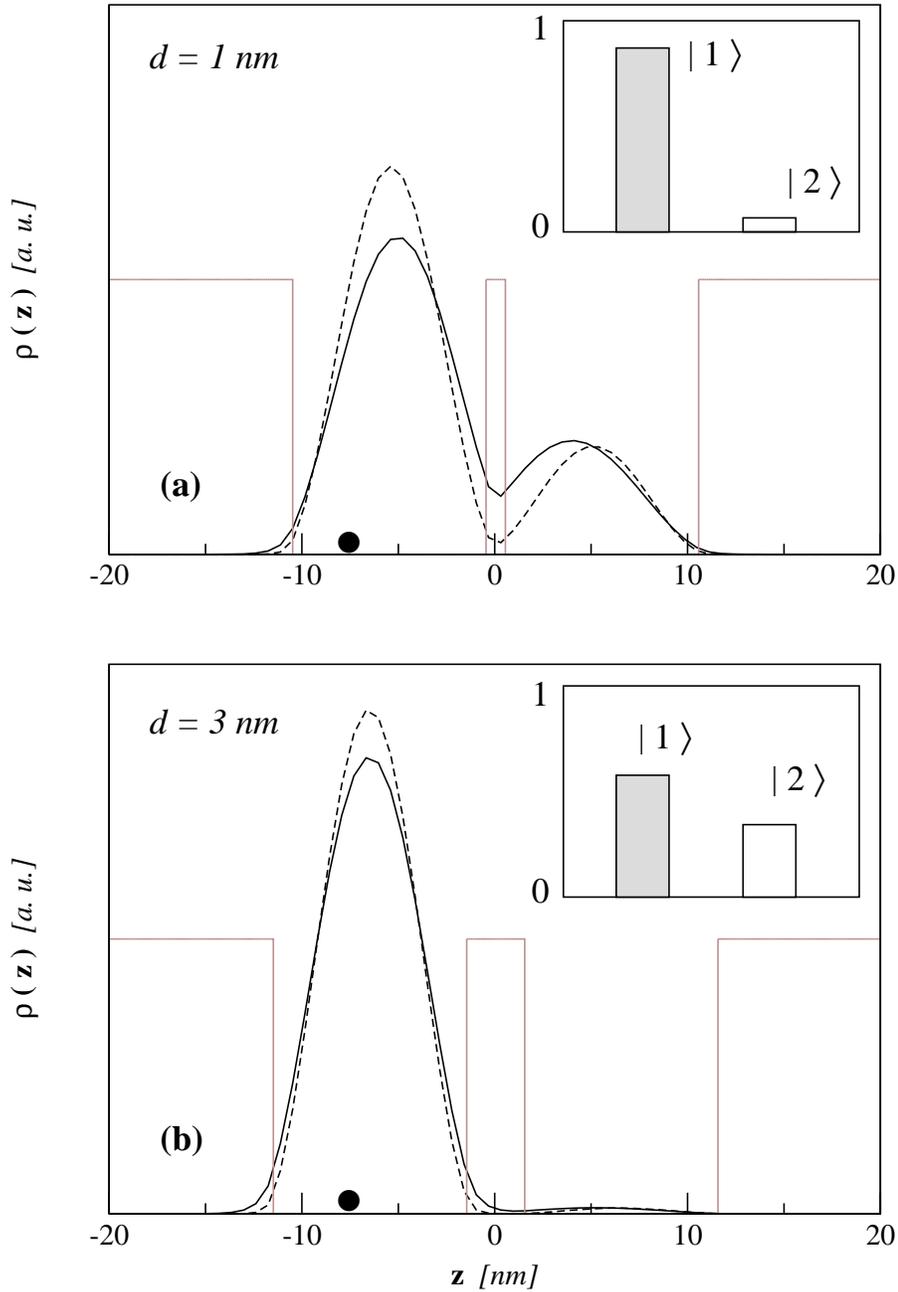}}
\caption{
Plot of the functions $ \rho ( z_{e} ) $ (continuous line) and
$ \rho ( z_{h} ) $ (dashed line), computed for the excitonic ground states
at $ d = 1 $ nm  (a) and $ d = 3 $ nm (b).
The definition of $\rho(z)$ and the positions of the fixed particles
(either electron or hole) are as in Fig.~\ref{scheme}.
The columns in the insets represent the square moduli
of the coefficients associated with the states
$ | 1 \rangle \equiv | \sigma_{g} \uparrow , 
\sigma_{g} \downarrow \rangle $
(grey column) and
$ | 2 \rangle \equiv | \sigma_{u} \uparrow , 
\sigma_{u} \downarrow \rangle $
(white column): the excitonic ground states are given,
to a good degree of approximation, 
by the superposition of these two states. At
$ d = 1$ nm   $ | c_{1} |^{2}=0.871 $ and $ | c_{2} |^{2}=0.066 $,
while at
$ d = 3$  nm  $ | c_{1} |^{2}=0.577 $ and $ | c_{2} |^{2}=0.344 $;
minor contributions arise from occupations
of higher-energetic single-particle states.
\label{exciton}
}
\end{figure}

We next consider systems of interacting carriers composed of
an equal number of electrons and holes. Let us start by
considering a single electron-hole pair 
(exciton) and the way in which its
ground state depends
on the width of the barrier.
As the inter-dot distance $d$ increases the splitting
between the energies of the bonding ($ \sigma_{g} $) and anti-bonding
($ \sigma_{u} $) states decreases both for electrons and for holes.
The energetic cost associated to the promotion of the two particles
from the bonding to the anti-bonding states becomes smaller and
comparable to the gain
in Coulomb energy arising from the 
correlation of their spatial distributions
along $z$. In Fig.~\ref{exciton} we plot the
functions $ \rho ( z )$ of electrons and
holes, associated to given positions of the other carrier,
at inter-dot distances
of $ d = 1 $ nm (a) and $ d = 3 $ nm (b); the two insets represent the
contributions to the electron-hole ground states of
$ | 1 \rangle \equiv | \sigma_{g}^{e} \uparrow , \sigma_{g}^{h}
\uparrow \rangle $
and
$ | 2 \rangle \equiv | \sigma_{u}^{e} \uparrow ,
\sigma_{u}^{h} \uparrow \rangle $.
The decreasing tunneling goes with an increasing correlation
(electron more localized around 
the hole and vice versa) and an increasing
contribution from the $ | 2 \rangle $ state.
The slight difference between the plots 
associated to electrons and holes at
each inter-dot distance are due to 
the differences in the barrier heights
(400 meV for electrons and 215 meV for holes) and in the effective
masses ($ m_{e}^{*} = 0.067\: m_{0} $ and $ m_{h}^{*} = 0.38\: m_{0} $)
of the two carriers: as a consequence 
electrons tunnel more than holes and tend
to be less localized within one QD.
It is worth noticing that at $ d = 3 $ nm 
the electronic tunneling still induces
a pronounced splitting between 
the two delocalized bonding and anti-bonding
single-particle states ($ \epsilon_{b}= 35.11$ meV and
$ \epsilon_{a}= 37.52$ meV). In spite of this, 
due to Coulomb correlation and
to the reduced tunneling of holes, the energetic value and the spatial
distribution of the excitonic ground state closely resembles that of an
exciton in a single QD; besides, the splitting between the ground and
the first excited states is negligible \cite{insertioV}.
In other words the ``excitonic tunneling'' is
suppressed at smaller inter-dot distances than the electronic one.

\begin{figure}
\centerline{\includegraphics[width=0.85\columnwidth]{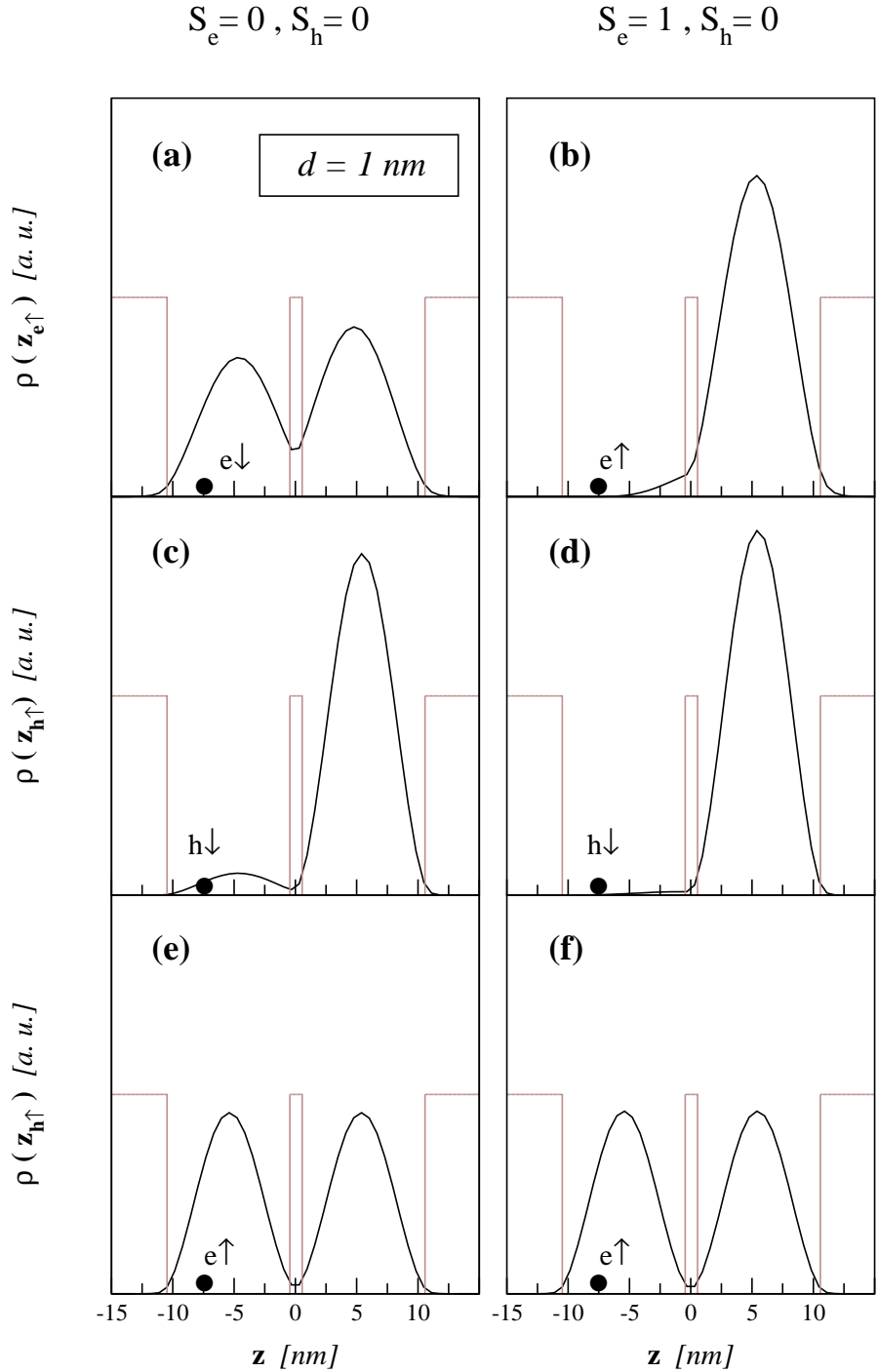}}
\caption{
The two spin configurations $ (S_{e}=0,S_{h}=0) $ and $ (S_{e}=1,S_{h}=0) $
are considered at $d=1$ nm.
The plots represent the spatial distribution of a carrier
of spin orientation as specified by the subscript
of the z-coordinate, given the
position of another carrier is fixed (whose
type and spin orientation
are drawn close to the corresponding black circle).
The position of the fixed particles 
is the one adopted in Fig.~\ref{scheme}.
\label{f0010d1}
}
\end{figure}

\begin{figure}
\centerline{\includegraphics[width=0.85\columnwidth]{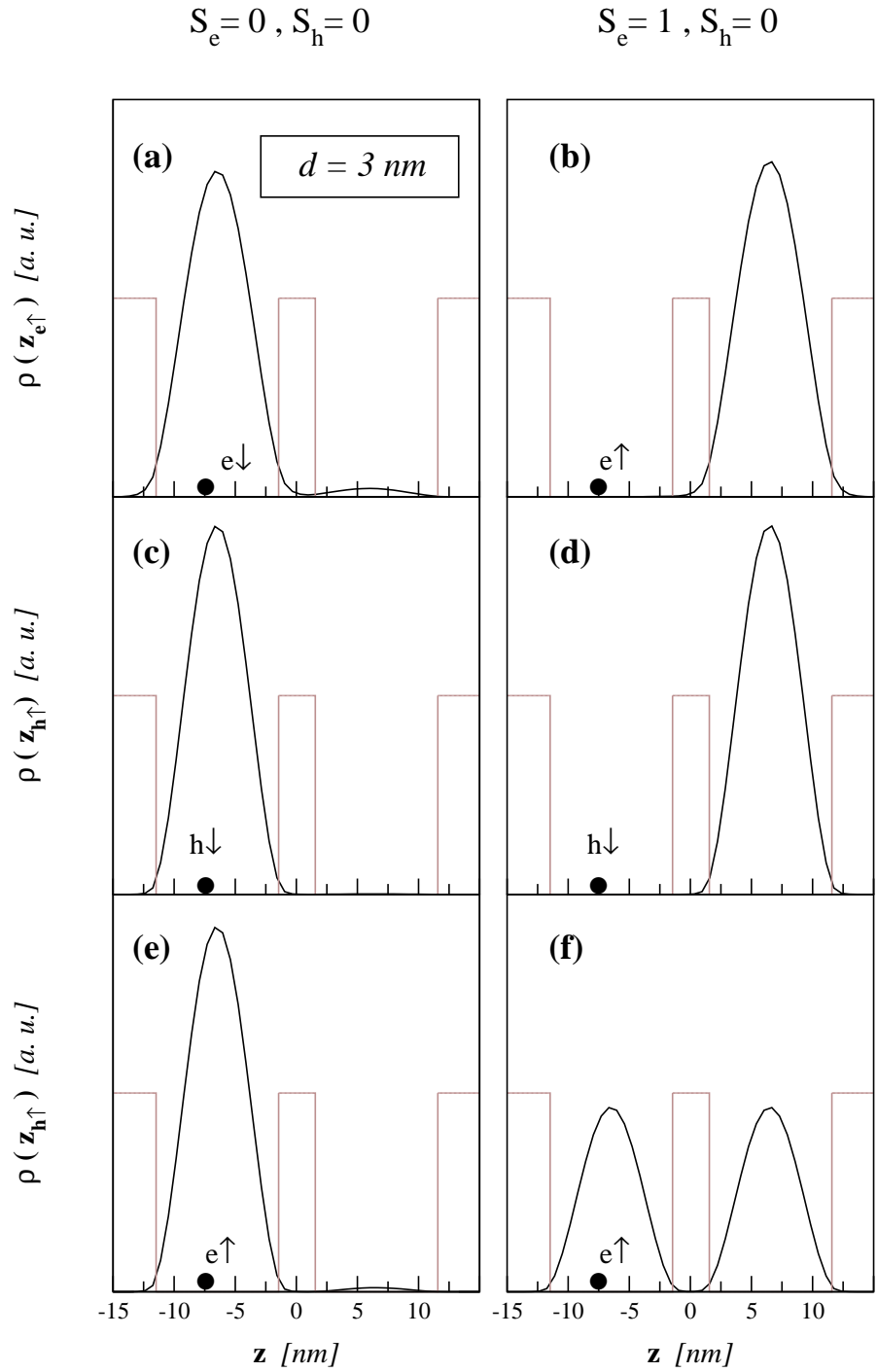}}
\caption{
Same functions as in Fig.~\ref{f0010d1}, but for
an inter-dot distance of 3 nm.
\label{f0010d3}
}
\end{figure}

If the double dot is occupied by two electrons and two holes, 
both attractive and
repulsive interactions are present. Intuitively one would expect
carriers with identical charge to avoid each other
and carriers of opposite charge to look for each other: the
interplay between such trends is directed by the
values of $ d $ and by those of $ S_{e} $ and $ S_{h} $.
In Fig.~\ref{f0010d1} we compare such correlations
for two different values of the
electron and hole spin quantum numbers and for $ d = 1 $ nm.
The singlet-singlet lowest state
($ S_{e} = 0, S_{h} = 0 $) is characterized by a small
correlation between the electrons, (a), 
and by a more pronounced one between
holes, (c). Analyzing the eigenfunction 
associated to this state, one
observes that approximately only the electron 
single-particle state $ \sigma_{g}^{e} $
is (twice) occupied, while for holes strong 
contributions arise from both bonding
and anti-bonding states. As already mentioned, this difference
between the behaviors of the two 
carriers depends on the fact that a gain
in Coulomb energy has a greater 
kinetic cost for electrons than for holes.
Besides, the spatial distribution of holes, (e),
is not affected by the position of
the electrons \cite{insertioVI}.
Let us now compare these correlation functions
with the corresponding ones associated to
the triplet-singlet configuration ($ S_{e} = 1, S_{h} = 0 $).
Again the correlation among electrons and holes is
negligible, (f): the two electrons and the two holes
subsystems can thus be understood to a good extent
independently from each other. As in the case of the prototypical
state $ | 3 \rangle $, that the state of the two 
electrons here resembles,
the probability of finding two electrons 
in the same QD is negligible, (b).
Such spatial separation of the electrons induces a more
pronounced separation for the holes too, as shown by the
flattening of the smallest peak in the left well
[see Fig.~\ref{f0010d1} (b), (d)].
The differences between the two spin configurations are even
more dramatic at bigger inter-dot distances. In
Fig.~\ref{f0010d3} we show the same
correlation functions at $ d = 3 $ nm.
The triplet-singlet configuration [Figs.~\ref{f0010d1} (b),
(d), (f)]
shows the same features already observed at $ d = 1 $ nm; 
here the two electrons (holes) are perfectly localized 
in different QDs, due to the suppressed tunneling.
The singlet-singlet configuration [Figs.~\ref{f0010d1}
(a), (c), (e)] instead has undergone
a transition to a phase in which all carriers are localized in either QD
(due to symmetry). If the position of one of the four particles is fixed
in one QD, all the others are localized in the same one.
This somehow surprising effect can be explained in the
following way: in a mean
field picture, due to the substantial similarity of
the electron and hole wavefunctions, the localization of
the two excitons in two different QDs or in the same
one makes no difference with respect to the Coulomb energy
because there is cancellation of terms of opposite sign.
If correlation comes into play, however, the localization
of all particles in the same QD gives rise
to the so-called ``biexcitonic binding energy'' $\Delta E$
(which is defined as the difference between twice the
energy of the excitonic ground state and that of the biexcitonic one).
Specifically, the binding energy $\Delta E$ is 
due to the correlations among
the $x$ and $y$ directions: as in the case of the two electrons in
Fig.~\ref{fthree}, such correlations become strongly
effective and lower the Coulomb energy when particles are
localized in the same QD.

The comparison between the correlation 
functions corresponding to equal spin
configurations at different inter-dot distances
shows a trend similar to that  of the two electrons alone.
The population of the anti-bonding states
increases with decreasing bonding-antibonding splitting, thus allowing a
more pronounced spatial correlation between identical carriers.
Such dependence is particularly clear for the singlet states, while for the
triplet ones a high degree of correlation is already guaranteed by the
permutational symmetry of the few-particle wavefunction (i.e., by the
fermionic nature of electrons and holes).

\begin{figure}
\centerline{\includegraphics[width=0.85\columnwidth]{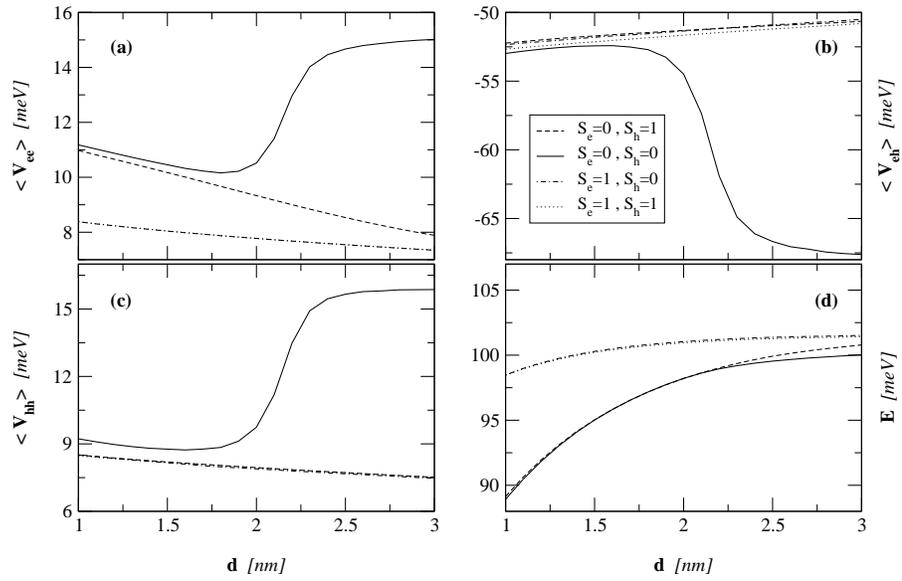}}
\caption{
Plots (a), (b)
and (c) represent the $d$-dependence of the average 
values of the different contributions to
the Coulomb energy for the eigenstates 
of lowest total energy associated
to each of the four spin configurations $ ( S_{e} , S_{h} ) $. 
The curves associated to the $ (1,0) $ and
$ (1,1) $ spin arrangements are identical
in (a), and, together with $ (0,1) $, in (c).
In (d) we plot the total energies.
\label{contr}
}
\end{figure}

The described behaviours are reflected in
the values of the different contributions to
the mean Coulomb energies of each spin configuration
(Fig.~\ref{contr}).
Let us start by considering the 
three spin arrangements $ ( S_{e} , S_{h} ) =
(1,1),(0,1),(1,0)$. The contributions to 
the Coulomb energy associated to
the electron-electron interaction (a) all monotonically decrease with
increasing $d$: the two electrons, each in a different QD, get 
more and more far apart. If $ S_{e}=1 $ the spatial 
separation of the two electrons is a direct consequence 
of the permutational symmetry of the wavefunction; if
$ S_{e}=0 $ the same effect arises from a proper linear combination of
different Slater determinants and from the corresponding 
occupation of the electronic $ \sigma_{u}^{e} $ 
orbitals (which in turn depends on the tunneling:
from which the different slopes of the $ S_{e}=0 $ and $ S_{e}=1 $ curves).
Analogous behaviours are seen to occur with respect to the hole-hole
interaction (c). The main difference as compared to 
the previous case is the
higher degree of correlation among such carriers in the $ S_{h} = 0 $ case,
already at small inter-dot distances.
The trends of the electron-hole Coulomb (b)
interactions are hardly distinguishable one from the other.
The monotonic decrease (in modulus) of
$ \langle V_{eh} \rangle $ reflects that of the interaction energy among
carriers localized in different QDs when they get far apart.
The plots associated to the singlet-singlet 
configuration (continuous lines), however,
show a transition towards a phase in which all carriers are localized in
a same QD, already put in evidence in Fig.~\ref{f0010d3}.
 The absolute values of all Coulomb terms correspondingly go through
an abrupt increase for values of $d$ in the range between 2 and 2.5 nm.
Let us finally consider the total energies (d).
The lowest singlet-singlet state turns out to
be the system's ground state at any inter-dot distance. For
$ d \lesssim 2$ nm the $ S_{e} = 0 $ and $ S_{e} = 1 $ 
configurations are degenerate with
respect to the value of $ S_{h} $; 
the difference between the total energies
follows that between the two-electrons triplet and singlet states. As $d$
increases, the energies of all spin configurations but
$S_{e}=S_{h}=0$ asymptotically
tend to a value which is twice the energy of
the excitonic ground state in a single QD.
The energy of the singlet-singlet state instead tends to that of
the biexcitonic ground state in a single QD:
the difference between these two asymptotic values
is the already mentioned biexcitonic binding energy $\Delta E$.

\section{Summary}
We have presented a unified theoretical description of 
the many-body states of a few electrons (holes) or few 
excitons confined in coupled quantum dots.
In these systems, interdot coupling controls the competition 
between kinetic energy and Coulomb interactions, and can reach
regimes far out of those accessible in natural molecules. 
The resulting ground state is therefore very different
for different values of the coupling: We have
shown that a system of few electrons is characterized 
by different spin configurations depending on the inter-dot
coupling, and we have extensively discussed the marked variations 
arising in the electron-electron correlation functions. 
In the case of two electrons and two holes we have identified
the ground state corresponding to both pairs localized in one 
of the dots (weak coupling) or distributed in both dots 
(strong coupling). Tuning such phases by external fields is 
possible, and is found to induce novel quantum effects that 
will be described elesewhere \cite{tobepubl}.

Manifestations of transitions between such phases in addition 
or optical spectra are expected to lead to a direct experimental 
verification of many-body-theory predictions, and to the
experimental control of the many-body states in nanoscale 
devices. 

\section{Acknowledgements} This paper was supported in part by the
EC through the SQID and Ultrafast Projects, and by INFM
through PRA SSQI.



\begin{thebibliography}{90}

\bibitem{book1}
L.~Jacak, P.~Hawrylak, and A.~W\'ojs,
{\it Quantum Dots,} (Springer, Berlin, 1998).

\bibitem{book2}
D.~Bimberg, M.~Grundmann, and N.~N.~Ledentsov, 
{\it Quantum Dot Heterostructures,}
(Wiley, Chichester, 1998).

\bibitem{book3}
U.~Woggon, {\it Optical Properties of Semiconductor Quantum Dots,}
(Springer, Berlin, 1997).

\bibitem{artificialatoms} M.~A.~Kastner, Phys.~Today {\bf 46,} 24 (1993).

\bibitem{lloyd}
S.~Lloyd, Science {\bf 261,} 1569 (1993).

\bibitem{review}
For a review see
L.~P.~Kouwenhoven, C.~M.~Marcus, P.~L.~McEuen,
S.~Tarucha, R.~M.~Westervelt, and N.~S.~Wingreen, in
{\em Mesoscopic Electron Transport,} edited by L.~L.~Sohn,
L.~P.~Kouwenhoven, and G.~Schoen, (Kluwer, Dordrecht, 1997), p.~105.

\bibitem{leoscience}
Leo Kouwenhoven, Science {\bf 268,} 1440 (1995).

\bibitem{leocrystal}
L.~P.~Kouwenhoven, F.~W.~J.~Hekking, B.~J.~van Wees, C.~J.~P.~M.~Harmans,
C.~E.~Timmering, and C.~T.~Foxon, Phys.~Rev.~Lett.~{\bf 65,} 361 (1990).

\bibitem{planarlinear}
F.~R.~Waugh, M.~J.~Berry, D.~J.~Mar, R.~M.~Westervelt, K.~L.~Campman,
and A.~C.~Gossard, Phys.~Rev.~Lett.~{\bf 75,} 705 (1995); F.~R.~Waugh,
M.~J.~Berry, C.~H.~Crouch, C.~Livermore, D.~J.~Mar, R.~M.~Westervelt,
K.~L.~Campman, and A.~C.~Gossard, Phys.~Rev.~B {\bf 53,} 1413 (1996);
C.~Livermore, C.~H.~Crouch, R.~M.~Westervelt, K.~L.~Campman, and
A.~C.~Gossard, Science {\bf 274,} 1332 (1996); 
T.~H.~Wang and S.~Tarucha, Appl.~Phys.~Lett.~{\bf 71,} 2499 (1997);
A.~S.~Adourian,
C.~Livermore, R.~M.~Westervelt, K.~L.~Campman, and A.~C.~Gossard,
Appl.~Phys.~Lett.~{\bf 75,} 424 (1999).

\bibitem{planarnonlinear}
C.~H.~Crouch, C.~Livermore, R.~M.~Westervelt,
K.~L.~Campman, and A.~C.~Gossard, Appl.~Phys.~Lett.~{\bf 71,} 817 (1997).

\bibitem{planarth}
K.~A.~Matveev, L.~I.~Glazman, and H.~U.~Baranger, Phys.~Rev.~B {\bf 53,}
1034 (1996); {\em ibid.}~{\bf 54,} 5637 (1996);
John M.~Golden and Bertrand I.~Halperin, 
Phys.~Rev.~B {\bf 53,} 3893 (1996);
{\em ibid.}~{\bf 54,} 16757 (1996).

\bibitem{Hubbardsimple}
C.~A.~Stafford and S.~Das Sarma, Phys.~Rev.~Lett.~{\bf 72,} 3590 (1994);
Gerhard Klimeck, Guanlong Chen, and Supriyo Datta, Phys.~Rev.~B {\bf 50,}
2316 (1994); Guanlong Chen, Gerhard Klimeck, Supriyo Datta,
Guanhua Chen, and William A.~Goddard III, 
Phys.~Rev.~B {\bf 50,} 8035 (1994);
R.~Kotlyar and S.~Das Sarma, Phys.~Rev.~B {\bf 56,} 13235 (1997).

\bibitem{blickI}
R.~H.~Blick, R.~J.~Haug, J.~Weis, D.~Pfannkuche, K.~von Klitzing,
and K.~Eberl, Phys.~Rev.~B {\bf 53,} 7899 (1996); R.~H.~Blick,
D.~Pfannkuche, R.~J.~Haug, K.~von Klitzing, and K.~Eberl,
Phys.~Rev.~Lett.~{\bf 80,} 4032 (1998).

\bibitem{blickII}
Robert H.~Blick, Daniel W.~van der Weide, Rolf J.~Haug, and Karl Eber,
Phys.~Rev.~Lett.~{\bf 81,} 689 (1998).

\bibitem{PAT}
T.~H.~Oosterkamp, T.~Fujisawa, W.~G.~van der Wiel, K.~Ishibashi,
R.~V.~Hijman, S.~Tarucha, and L.~P.~Kouwenhoven, 
Nature (London) {\bf 395,} 873 (1998). 

\bibitem{spontaneous}
Toshimasa Fujisawa, Tjerk H.~Oosterkamp, Wilfred G.~van der Wiel, Benno
W.~Broer, Ram\'on Aguado, Seigo Tarucha, and Leo P.~Kouwenhoven,
Science {\bf 282,} 932 (1998).

\bibitem{vaart}
N.~C.~van der Vaart, S.~F.~Godijn, Y.~V.~Nazarov, C.~J.~P.~M.~Harmans,
J.~E.~Mooij, L.~W.~Molenkamp, and C.~T.~Foxon, Phys.~Rev.~Lett.~{\bf 74,}
4702 (1995).

\bibitem{magnetization}
T.~H.~Oosterkamp, S.~F.~Godijn, M.~J.~Uilenreef, Y.~V.~Nazarov,
N.~C.~van der Vaart, and Leo P.~Kouwenhoven, Phys.~Rev.~Lett.~{\bf 80,}
4951 (1998).

\bibitem{bunching}
M.~Brodsky, N.~B.~Zhitenev, R.~C.~Ashoori, L.~N.~Pfeiffer, and K.~W.~West,
Phys.~Rev.~Lett.~{\bf 85,} 2356 (2000).

\bibitem{schmidt}
T.~Schmidt, R.~J.~Haug, K.~von Klitzing, A.~F\"orster, and H.~L\"uth,
Phys.~Rev.~Lett.~{\bf 78,} 1544 (1997).

\bibitem{guy}
David Guy Austing, Takashi Honda, and Seigo Tarucha,
Jpn.~J.~Appl.~Phys.~{\bf 36}, 1667 (1997); D.~G.~Austing, T.~Honda,
K.~Muraki, Y.~Tokura, and S.~Tarucha,
Physica B {\bf 249-251,} 206 (1998).

\bibitem{bryant}
Garnett W.~Bryant, Phys.~Rev.~B {\bf 48,} 8024 (1993).

\bibitem{oh}
J.~H.~Oh, K.~J.~Chang, G.~Ihm, and S.~J.~Lee, Phys.~Rev.~B {\bf 53,}
R13264 (1996).

\bibitem{tamura}
Hiroyuki Tamura, Physica B {\bf 249-251,} 210 (1998).

\bibitem{tokura}
Y.~Tokura, D.~G.~Austing, and S.~Tarucha, in {\em Proceedings of the 24th
International Conference on the Physics of
Semiconductors,} edited by D.~Gershoni, (World Scientific, on
CD-ROM, 1999); J.~Phys.: Condens.~Matter {\bf 11,} 6023 (1999);
Y~Tokura, S.~Sasaki, D.~G.~Austing, and S.~Tarucha, Physica E
{\bf 6,} 676 (2000).

\bibitem{ssc}
M.~Rontani, F.~Rossi, F.~Manghi, and E.~Molinari,
Solid State Comm.~{\bf 112,} 151 (1999);
Mat.~Res.~Soc.~Symp.~Proc.~{\bf 571,} 179 (2000).

\bibitem{asano}
Yasuhiro Asano, Phys.~Rev.~B {\bf 58,} 1414 (1998).

\bibitem{bart}
B.~Partoens and F.~M.~Peeters, Phys.~Rev.~Lett.~{\bf 84,} 4433 (2000);
Mart\'{\i} Pi, Agust\'{\i}
Emperador, Manuel Barranco, and Francesca Garcias, Phys.~Rev.~B
{\bf 63,} 115316 (2001).

\bibitem{palacios}
Juan Jos\'e Palacios and Pawel Hawrylak, Phys.~Rev.~B {\bf 51,}
1769 (1995).

\bibitem{dagotto}
Jun Hu, E.~Dagotto, and A.~H.~MacDonald, Phys.~Rev.~B {\bf 54,}
8616 (1996).

\bibitem{aoki}
Hiroshi Imamura, Peter A.~Maksym, and Hideo Aoki, Phys.~Rev.~B {\bf 53,}
12613 (1996); Hiroshi Imamura, Hideo Aoki, and Peter A.~Maksym,
{\em ibid.}~{\bf 57,} R4257 (1998); Hiroshi Imamura,
Peter A.~Maksym, and Hideo Aoki, {\em ibid.}~{\bf 59,} 5817 (1999).

\bibitem{tejedor}
L.~Mart\'{\i}n-Moreno, L.~Brey, and C.~Tejedor,
Phys.~Rev.~B {\bf 62,} R10633 (2000).

\bibitem{mayrock}
O.~Mayrock, S.~A.~Mikhailov, T.~Darnhofer, and U.~R\"ossler,
Phys.~Rev.~B {\bf 56,} 15760 (1997);
B.~Partoens, A.~Matulis, and F.~M.~Peeters, {\em ibid.}~{\bf 57,}
13039 (1998).

\bibitem{peeters}
B.~Partoens, V.~A.~Schweigert, and F.~M.~Peeters,
Phys.~Rev.~Lett.~{\bf 79,} 3990 (1997).

\bibitem{yannouleas}
Constantine Yannouleas and Uzi Landman, Phys.~Rev.~Lett.~{\bf 82,}
5325 (1999).

\bibitem{helium}
B.~Partoens, A.~Matulis, and F.~M.~Peeters, Phys.~Rev.~B {\bf 59,}
1617 (1999).

\bibitem{eto}
Mikio Eto, Solid-State Electron.~{\bf 42,} 1373 (1998).

\bibitem{natalia}
Yu.~E.~Lozovik and N.~E.~Kaputnika, Physica Scripta {\bf 57,}
542 (1998); N.~E.~Kaputnika and Yu.~E.~Lozovik,
Fiz.~Tverd.~Tela {\bf 40,} 2127 (1998)
[Sov.~Phys.~Solid State {\bf 40,} 1929 (1998)].

\bibitem{leburton}
Satyadev Nagaraja, Jean-Pierre Leburton, and Richard M.~Martin,
Phys.~Rev.~B {\bf 60,} 8759 (1999); Andreas Wensauer, Oliver
Steffens, Michael Suhrke, and Ulrich R\"ossler,
Phys.~Rev.~B {\bf 62,} 2605 (2000).

\bibitem{tapash}
Tapash Chakraborty, V.~Halonen, and P.~Pietil\"ainen,
Phys.~Rev.~B {\bf 43,} 14289 (1991).

\bibitem{molenkamp}
Xiaoshuang Chen, H.~Buhmann, and L.~W.~Molenkamp,
Phys.~Rev.~B {\bf 61,} 16801 (2000).


\bibitem{qc}
Guido Burkard, Daniel Loss, and David P.~DiVincenzo,
Phys.~Rev.~B {\bf 59,} 2070 (1999);
Xuedong Hu and S.~Das Sarma, Phys.~Rev.~A {\bf 61,} 62301 (2000).

\bibitem{Amaha} S.~Amaha, D.~G.~Austing, Y.~Tokura, K.~Muraki,
K.~Ono, and S.~Tarucha, Solid State Comm.~(2001), this issue. 

\bibitem{schedelbeck:97}
G.~Schedelbeck, W.~Wegschreider, M.~Bichler, G.~Abstreiter,
Science {\bf 278,} 1792 (1997).

\bibitem{fafard:00}
S.~Fafard, M.~Spanner, J.~P.~McCaffrey, and Z.~R.~Wasilewski,
Appl.~Phys.~Lett.~{\bf 75,} 2268 (2000), and references therein.

\bibitem{bayer:01}
M.~Bayer, P.~Hawrylak, K.~Hinzer, S.~Fafard,
M.~Korkusinski, Z.~R.~Wasilewski, O.~Stern,
A.~Forchel, Science {\bf 291,} 451 (2001).

\bibitem{insertio}
For single quantum dots it is now known from
previous theoretical and experimental work that few-particle Coulomb
correlations dominate the optical spectra in the non-linear regime.
See, e.g., Refs.~\cite{hartmann:00,zrenner:00}, and references therein.

\bibitem{hartmann:00}
A.~Hartmann, Y.~Ducommun, E.~Kapon, U.~Hohenester, and E.~Molinari,
Phys.~Rev.~Lett.~{\bf 84,} 5648 (2000).

\bibitem{zrenner:00}
A.~Zrenner, J.~Chem.~Phys.~{\bf 112,} 7790 (2000).

\bibitem{filippoprl}
F.~Troiani, U.~Hohenester, and E.~Molinari, unpublished.

\bibitem{bennet:00}
C.~H.~Bennet and D.~P.~DiVincenzo, Nature (London) {\bf 404,} 247 (2000).

\bibitem{troiani:00}
F.~Troiani, U.~Hohenester, and E.~Molinari, Phys.~Rev.~B
{\bf 62,} R2263 (2000), and references therein.

\bibitem{biolatti:00}
E.~Biolatti, R.~C.~Iotti, P.~Zanardi, and F.~Rossi,
Phys.~Rev.~Lett.~{\bf  85,} 5647 (2000).

\bibitem{slater}
John C.~Slater, {\em Quantum Theory of Molecules
and Solids,} vol.~1, (McGraw-Hill, New York, 1963).

\bibitem{insertioII}
The notation slightly differs from that used in
Ref.~\cite{ssc}, where $\pm$ refers to the reflection with respect to
the $xy$ plane. Reference \cite{tejedor} defines this
plane reflection as ``parity.''

\bibitem{capacitance}
R.~C.~Ashoori, Nature (London) {\bf 379,} 413 (1996).

\bibitem{tarucha}
S.~Tarucha, D.~G.~Austing, T.~Honda, R.~J.~van der Hage,
and L.~P.~Kouwenhoven, Phys.~Rev.~Lett.~{\bf 77,} 3613 (1996).

\bibitem{leoexcited}
L.~P.~Kouwenhoven, T.~H.~Oosterkamp, M.~W.~S.~Danoesastro,
M.~Eto, D.~G.~Austing, T.~Honda, and S.~Tarucha, Science {\bf 278,}
1788 (1997).

\bibitem{spectroscopy}
C.~W.~J.~Beenakker, Phys.~Rev.~B {\bf 44,} 1646 (1991).

\bibitem{kohn}
W.~Kohn, Phys.~Rev.~{\bf 123,} 1242 (1961).

\bibitem{Raman}
D.~J.~Lockwood, P.~Hawrylak, P.~D.~Wang, C.~M.~Sotomayor Torres,
A.~Pinczuk, and B.~S.~Dennis, Phys.~Rev.~Lett.~{\bf 77,} 354 (1996);
C.~Sch\"uller, K.~Keller, G.~Biese, E.~Ulrichs, L.~Rolf, C.~Steinebach,
D.~Heitmann, and K.~Eberl, {\em ibid.}~{\bf 80,} 2673 (1998).

\bibitem{haug:93}
H.~Haug and S.~W.~Koch,  
{\em Quantum theory of the optical and
electronic properties of semiconductors,} (World Scientific, 
Singapore, 1993).

\bibitem{wignerth}
Daniel C.~Mattis, {\em The Theory of Magnetism,} (Harper, New York, 1965),
p.~91.

\bibitem{singlet}
The magnetic field breaks time-reversal symmetry and
induces singlet-triplet transitions. In the artificial
Helium this proceeds with increments
of the quantum of angular momentum $\hbar$ [M.~Wagner, U.~Merkt, and
A.~V.~Chaplik, Phys.~Rev.~B {\bf 45,} 1951 (1992)].

\bibitem{insertioIII}
Here by ``kinetic energy'' we mean the sum of the
single-particle contributions to the total energy, thus
including the effect of the external confinement potential.

\bibitem{insertioIV}
We describe a
configuration listing the occupied single-particle levels,
labeled as $nm_{g,u}^{\pm}$: $n$ is the radial quantum number,
$m$ assumes the symbols $\sigma,\pi,\delta,\ldots$ corresponding to the
azimuthal quantum number $m=0,1,2,\ldots$, the superscript + (-) stands
for positive (negative) values of $m$, and the subscript $g$ ($u$)
refers to even (odd) parity. Cf.~Ref.~\cite{slater}.

\bibitem{maksym}
P.~A.~Maksym, Phys.~Rev.~B {\bf 53,} 10871 (1996).

\bibitem{arpack}
R.~B.~Lehoucq, K.~Maschhoff, D.~C.~Sorensen, C.~Yang, ARPACK
computer code, {\tt http://www.caam.rice.edu/software/ARPACK/}.

\bibitem{landau}
L.~D.~Landau and E.~M.~Lifshitz, {\em Quantum
Mechanics -  Non Relativistic Theory,} (Pergamon Press, Oxford, 1958).

\bibitem{pfannkuche}
Daniela Pfannkuche, Vidar Gudmundsson,
and Peter A.~Maksym, Phys.~Rev.~B {\bf 47,} 2244 (1993).

\bibitem{prb}
M.~Rontani, F.~Rossi, F.~Manghi, and E.~Molinari, Phys.~Rev.~B {\bf 59,}
10165 (1999).

\bibitem{tobepubl}
M.~Rontani, G.~Goldoni, F.~Manghi, and E.~Molinari, unpublished.

\bibitem{insertioV}
The main contributions to the 
excitonic first excited state come from the states
$ | 3 \rangle \equiv | \sigma_{g}^{e} \uparrow ,
\sigma_{u}^{h} \uparrow \rangle $ and
$ | 4 \rangle \equiv | \sigma_{u}^{e} \uparrow ,
\sigma_{g}^{h} \uparrow \rangle $. 
At $ d = 1 $ nm $ | c_{3} |^{2} = 0.852$
and  $ | c_{4} |^{2} = 0.0836 $, while at $ d = 3 $ nm
$ | c_{3} |^{2} = 0.573$ and $ | c_{4} |^{2} = 0.348 $; 
in the latter case
the correlation function looks very much 
like that of the ground state.

\bibitem{insertioVI}
Formally this means that the
four-particle wavefunction can be written 
to a good degree of approximation
in a factorized form: $\psi ( \vec{r_{e\uparrow}} , 
\vec{r_{e\downarrow}} ,
\vec{r_{h\uparrow}} ,\vec{r_{h\downarrow}} )
\simeq \phi_{e} (\vec{r_{e\uparrow}}
 , \vec{r_{e\downarrow}}) \phi_{h} (\vec{r_{h\uparrow}},
\vec{r_{h\downarrow}})$.

\end{thebibliography}
\end{document}